\documentstyle[prl,aps,epsfig,multicol]{revtex}
\begin{document}
\def\be{\begin{equation}}
\def\ee{\end{equation}}
\def\bearr{\begin{eqnarray}}
\def\eearr{\end{eqnarray}}

\draft
\title{ Vortex formation in a fast rotating Bose-Einstein condensate}

\author{Tarun Kanti Ghosh}
\address
{ The Abdus Salam International Centre for Theoretical
Physics, Strada Costiera 11, 34100 Trieste, Italy.}

\date{\today}
\maketitle

\begin{abstract}
We study rotational motion of an interacting atomic Bose-Einstein condensate 
confined in a quadratic-plus-quartic potential. We calculate the lowest energy
surface mode frequency and show that a symmetric trapped (harmonic and quartic) 
Bose-Einstein condensate breaks the rotational symmetry of the Hamiltonian 
when rotational frequency is greater than one-half of the lowest energy surface mode 
frequency. We argue that the formation of a vortex is not possible in a non-interacting
as well as in an attractive Bose-Einstein condensate confined in a harmonic trap due to the 
absence of the spontaneous shape deformation, but it can occur which leads to the vortex formation 
if we add an additional quartic potential. Moreover, the spontaneous shape deformation and consequently 
the formation of a vortex in an attractive system depends on the strengths of the two-body 
interaction and the quartic potential.

\end{abstract}
\pacs{PACS numbers: 03.75.Lm, 05.30.Jp}   

\begin{multicols}{2}[]
\section{Introduction}
After the creation and realization of the Bose-Einstein condensate (BEC)
in an alkali atomic trapped Bose gas, a new field, namely strongly correlated
atomic system, has been opened up \cite{rmp}. It is always interesting to study 
the consequences of an external perturbation to the system. One way to perturb 
the system is by applying external rotation to the system. The rotation of a
macroscopic quantum fluid exhibits interesting counter-intuitive phenomena. For 
example, an atomic interacting(repulsive) BEC confined in a rotating harmonic 
trap produces quantized vortices for a sufficiently large rotation \cite{mad,mad1,abo}. 
The theoretical calculation for critical rotation frequency ($ \Omega_c $) based on purely 
thermodynamic arguments \cite{baym1} is significantly smaller than the observed value of 
$\Omega_c $. For sufficiently high stirring frequencies of the trap, the system
deforms spontaneously \cite{recati} to lower the energy barrier which prevents 
the nucleation of vortices at smaller $\Omega$ \cite{kramer} and this spontaneous
deformation is detected at the ENS experiment which leads to the vortex nucleation \cite{madison}. 
In fact, vortices are nucleated when the stirring potential with low anisotropies resonantly 
excites the quadrupole mode of the condensate, and induces large amplitudes oscillations of the
condensate, resulting in a dynamical instability \cite{castin}.

The harmonically trapped BEC becomes singular when the rotation frequency
is equal to or greater than the harmonic trap frequency, because the outward 
centrifugal force counteracts the inward force from the harmonic trap. 
We can eliminate the singularity at $ \Omega \geq 1 $ if 
we consider an additional stiffer radial potential (say, quartic potential for 
simplicity). There is a growing interest about the effect of an anharmonic potential on 
the properties of a rotating BEC \cite{fetter,lundh,ueda,baym,lundh1}. 
Recently, the quadratic-plus-quartic potential has been achieved experimentally at ENS
by superimposing a blue detuned laser beam to the magnetic harmonic trap \cite{fast}.

In spite of the fact that the $ \Omega_c $ in a harmonically trapped BEC
calculated from the thermodynamic arguments does not match with the
experimental values, many authors have studied $ \Omega_c $ in a BEC confined in
harmonic-plus-quartic trap based on the thermodynamic arguments \cite{lundh,ueda,baym,lundh1}.
In this work, we argue that a vortex formation is possible in an attractive
as well as non-interacting Bose gas confined in a quadratic-plus-quartic potential
and calculate the critical rotational frequencies based on the
spontaneous shape deformation of rotating Bose gas. 

This paper is organised as follows.
In section II, we write down the model Hamiltonian and variational macroscopic 
wave function, from which we calculate the variational energy. By minimizing the 
variational energy, we derive three coupled equations of the variational parameters.
In section III, we calculate the lowest energy surface mode frequency of an 
interacting BEC confined in a quadratic-plus-quartic potential.
In section IV, we discuss how spontaneous shape deformation transfer angular momentum and
vortex nucleation occurs in a harmonically 
trapped BEC. In section V, we discuss the 
possibility of the shape deformation and vortex nucleation of a 
BEC trapped in a quadratic-plus-quartic potential.
We present summary and conclusions of our work in section VI.

\section{The system and the Hamiltonian}
The equation of motion of the condensate wave function is described by the
mean-field Gross-Pitaevskii equation,
\be
i \hbar \frac{ \partial \psi (\vec r)}{\partial t} = [ - \frac{\hbar^2}{2m} \nabla^2
+ V(\vec r) + g |\psi(\vec r)|^2 - \Omega_0 L_z ] \psi(\vec r),
\ee
where  $ V(\vec r) = \frac{1}{2} m [(\omega_x^2 x^2 + \omega_y^2 y^2 + \omega_z^2
z^2) + \lambda \omega_0^2 (x^2 + y^2)^2 ]$ is the quadratic-plus-quartic potential with the small
anharmonic term $ \lambda $. Here, we assume $ \omega_x^2 = \omega_0^2 (1+ \epsilon) $ 
and $ \omega_x^2 = \omega_0^2 (1 -  \epsilon) $, 
where $ \epsilon = \frac{(\omega_x^2 - \omega_y^2)}{(\omega_x^2 + \omega_y^2)} $ is a very small 
deformation parameter due to the rotating deformed potential.
Also, $ g = \frac{4 \pi a \hbar^2}{m} $ is the strength of the mean-field interaction,
$ L_z = x p_y - y p_x $ is the $z$-component of the angular momentum operator, 
and $ \Omega_0 $ is the trap rotation frequency. When $ \epsilon = 0 $, it preserves the circular symmetry
of the system.

For simplicity, we shall study quasi-2D system.
The wave function in the $z$-direction is separable and is given by,
\be
\psi(z) = \frac{1}{(\sqrt{\pi} a_z)^{1/2}} e^{-\frac{z^2}{2 a_z^2}},
\ee
where $ a_z^2 = \frac{\hbar}{m \omega_z} $.
After integrating out the $z$-component, we get the effective 2D Gross-Pitaevskii
energy functional with the modified interaction strength 
$ g_2 = 2 \sqrt{2\pi}\hbar \omega_z a a_z $ \cite{ho}.  
One can write down the Lagrangian density corresponding to the quasi-2D system as follows:
\bearr \label{density}
{\cal L} & = & \frac{ i \hbar }{ 2 }(\psi\frac{\partial{\psi^{*}}}{\partial{ t }} -
\psi^{*}\frac{\partial\psi}{\partial t}) \\ \nonumber & + &
(\frac{\hbar^{2}}{2m} |\nabla \psi
|^{2} + V({x, y})|\psi |^{2} + \frac{g_2}{2} |\psi |^{4} - \Omega \psi^{*} L_z \psi ).
\eearr

Here, we use the time-dependent variational method \cite{perez} to study the properties
of a rotating BEC confined in a quadratic-plus-quartic potential.
In order to obtain the evolution of the condensate
we assume the most general Gaussian wave function,
\be \label{wave}
 \psi(X,Y,t) = C(t) e^{-\frac{1}{2}[ \alpha (t) X^2 + \beta (t) Y^2 - 2i \gamma
(t) X Y]},
\ee
where $ C(t) = \sqrt{\frac{ N \sqrt{D}}{\pi a_{0}^2 }} $ is the normalization constant.
and $ D = \alpha_{1} \beta_{1} - \gamma_{1}^2 $. 
The new co-ordinates $ X $ and $ Y $
are the dimensionless variables, $ X = \frac{x}{a_{0}}$, $ Y =
\frac{y}{a_{0}}$ where $ a_{0} = \sqrt{\frac{\hbar}{m\omega_{0}}} $ is the
oscillator length and $ 2 \omega_0^2 = \omega_x^2 + \omega_y^2 $ is the mean    
frequency.  Further, $ \alpha(t) = \alpha_1(t) + i \alpha_2(t) $,
$ \beta(t) = \beta_1(t) + i \beta_2(t) $ and $ \gamma(t) = \gamma_1(t) + i \gamma_2(t) $ are the 
time-dependent dimensionless complex variational parameters. The $ \alpha_1 $ and $ \beta_1 $ are
inverse square of the condensate widths in $ x $ and $ y $ directions,
respectively. 

We obtain the variational Lagrangian  $ L $ by substituting Eq. (\ref{wave})
into Eq. (\ref{density}) and integrating the Lagrangian density over the
space co-ordinates,
\bearr
\frac{L}{N\hbar\omega_{0}} & = & \nonumber
\frac{1}{4 D } [-(\beta_{1}\dot{\alpha_{2}}+\alpha_{1}\dot{\beta_{2}} - 2    
\gamma_1 \dot{\gamma_2}) + (\alpha_{1} + \beta_{1} ) D  
\\ \nonumber & + &   (\alpha_{2}^{2} + \gamma_{2}^2 ) \beta_{1}
 +  ( \beta_{2}^2 + \gamma_{2}^2 ) \alpha_{1} \\ \nonumber & - &  2 (
\alpha_{2} + \beta_{2} ) \gamma_{1} \gamma_{2}
 +  (\alpha_1 + \beta_{1}) + \epsilon( \beta_1 - \alpha_1)
\\ \nonumber & + & \frac{\lambda}{2} \frac{(3 \alpha_1^2 
+ 2 \alpha_1 \beta_1 + 3 \beta_1^2 + 4 \gamma_1^2)}{D} 
\\ & + & P D^{3/2}
 + 2 \Omega (\gamma_1(\beta_2 -\alpha_2)+ \gamma_2(\alpha_1
-\beta_1))],
\eearr
 where $ P = 2 \sqrt{\frac{2}{\pi}} \frac{(N-1) a}{a_{z}} $ is the dimensionless parameter
that indicates the strength of the mean-field interaction and it can be positive or negative depending on 
the sign of the $s$-wave scattering length $a$ \cite{fesh}.

The variational energy of the rotating condensate at equilibrium is given  in
terms of the inverse square width of the condensate
along the $x$ and $y$ directions and the phase parameter $ \delta \gamma_2$,
\bearr
\frac{ E}{N\hbar \omega_0} &  = & \nonumber \frac{1}{4} [ (\alpha_1 +
\beta_1  ) + \gamma_2^2 (\frac{1}{\alpha_1 } + \frac{1}{\beta_1 })
+ (\frac{1}{\alpha_1 } + \frac{1}{\beta_1 })
\\ \nonumber & + & \epsilon (\frac{1}{\alpha_1 } - \frac{1}{\beta_1 })
+\frac{\lambda}{2}(\frac{3}{\alpha_1^2}+ \frac{2}{\alpha_1 \beta_1} +\frac{3}{\beta_1^2})
\\  & + &  P \sqrt{\alpha_1  \beta_1 } + 2 \Omega \gamma_2
(\frac{1}{ \beta_1 } - \frac{1}{\alpha_1 })],
\eearr
where $ \Omega = \Omega_0/ \omega_0 $.
We have used the fact that $ \alpha_2 = \beta_2 = \gamma_1 =0 $, and also
$ \gamma_2 = \delta \gamma_2 $ to obtain the above equation.

One can get the equilibrium value of the variational parameters, $
\alpha_{10} $, $ \beta_{10} $ and $ \gamma_{20} $ by minimizing the energy
with respect to the variational parameters,

\be \label{c1}
\alpha_{10}^2(1 + \frac{P}{2} \sqrt{\frac{ \beta_{10}}{\alpha_{10}}}) =
\gamma_{20}^2 - 2 \Omega \gamma_{20} + 1 + \epsilon + \lambda (\frac{3}{\alpha_{10}}+ \frac{1}{\beta_{10}}),
\ee

\be \label{c2}
\beta_{10}^2(1 + \frac{P}{2}  \sqrt{\frac{\alpha_{10}}{\beta_{10}}}) =
\gamma_{20}^2 + 2 \Omega \gamma_{20} + 1 - \epsilon +  \lambda (\frac{1}{\alpha_{10}}+ \frac{3}{\beta_{10}}),
\ee

and
\be \label {c3}
\gamma_{20} = \Omega \frac{(\beta_{10} - \alpha_{10})}{(\beta_{10} + \alpha_{10})}.
\ee
The Eqs. (\ref{c1}), (\ref{c2}) and (\ref{c3}) describes how the shape of the condensate
changes due to the rotation.
The relation between the phase parameter and the variational widths is also obtained
in \cite{recati,sinha}.
The above equation can be re-written as
\be
\eta = \frac{<x^2>}{<y^2>} = \frac{\beta_{10}}{\alpha_{10}} = \frac{(\Omega + \gamma_{20})}{(\Omega -
\gamma_{20})}.
\ee
This implies that $ \Omega > \gamma_{20} $ since $ \eta $ is always positive.

The average angular momentum per particle is given by,
\be \label{ang}
\frac{<L_z>}{N \hbar} = \frac{1}{2}  \gamma_{20} (\frac{1}{\beta_{10}} -
\frac{1}{\alpha_{10}}) = \frac{\Omega}{2}\frac{ (1-\eta)^2}{\eta (\alpha_{10} +
\beta_{10})}.
\ee
This relation explicitly shows how the angular momentum transfers to a trapped
BEC due to the spontaneous shape deformation and consequently
how the vortices appears in the rotating BEC by calculating the average angular
momentum per particle \cite{dalfovo}. We would like to estimate the critical angular frequency 
from Eq.(\ref{ang}) by using the constraint $ <L_z> = N \hbar $. 

\section{lowest energy surface mode frequency}
To calculate the quadrupole mode frequency of BEC confined in a quadratic-plus-quartic potential,
we expand the Lagrangian in the following way: $ \alpha = \alpha_{10} + \delta \alpha_1 $,
$ \beta = \beta_{10} + \delta \beta_1 $, and $ \gamma = \delta \gamma_1 $. We keep only the
second order deviations from their equilibrium values.
Then the Lagrangian quadratic in the deviations takes the following form:

\bearr
L & = & \nonumber \frac{1}{4}[\frac{\delta \alpha_1 \dot{\delta \alpha_2}}{\alpha_{10}^2} +
\frac{\delta \beta_1 \dot{\delta \beta_2}}{\beta_{10}^2}+
2 \frac{\delta \gamma_1 \dot{\delta \gamma_2}}{\alpha_{10} \beta_{10}}
\\ \nonumber & + &
\frac{\delta \alpha_2^2 + \delta \gamma_2^2}{\alpha_{10}} +
\frac{\delta \beta_2^2 + \delta \gamma_2^2}{\beta_{10}}
\\ \nonumber & + &
\frac{ \alpha_{10} + \beta_{10}}{\alpha_{10} \beta_{10}}
(\frac{\delta \alpha_1^2}{\alpha_{10}^2}
+ \frac{\delta \beta_1^2}{\beta_{10}^2} + \frac{ \delta \alpha_1 \delta
\beta_1}{\alpha_{10}
\beta_{10}} + \frac{\delta \gamma_1^2}{\alpha_{10}\beta_{10}})
\\ \nonumber & - &
\frac{(\delta \alpha_1 + \delta \beta_1)}{ \alpha_{10} \beta_{10}} (\frac{\delta
\alpha_1}{\alpha_{10}} + \frac{\delta
\beta_1}{\beta_{10}})
\\ \nonumber & + &
\frac{\epsilon ( \beta_{10} - \alpha_{10})}{\alpha_{10}\beta_{10}}(\frac{\delta
\alpha_1^2}{\alpha_{10}^2}
+ \frac{\delta \beta_1^2}{\beta_{10}^2} + \frac{ \delta \alpha_1 \delta
\beta_1}{\alpha_{10}
\beta_{10}} + \frac{\delta \gamma_1^2}{\alpha_{10}\beta_{10}})
\\ \nonumber & - &
\frac{\epsilon (\delta \beta_1- \delta \alpha_1)}{\alpha_{10} \beta_{10}} (\frac{\delta 
\alpha_1}{\alpha_{10}} +
\frac{\delta \beta_1}{\beta_{10}}) 
\\ \nonumber & + & 
\frac{\lambda}{2(\alpha_{10}\beta_{10})^2}((9 \frac{\beta_{10}^2}{\alpha_{10}^2} + 
2 \frac{\beta_{10}}{\alpha_{10}}) \delta \alpha_1^2 + [(9 \frac{\alpha_{10}^2}{\beta_{10}^2} +
2 \frac{\alpha_{10}}{\beta_{10}}) \delta \beta_1^2 
\\ \nonumber & + &
 2 \delta \alpha_1 \delta \beta_1
+ 2 (4+ 3 \frac{\alpha_{10}}{\beta_{10}} + 3 \frac{\beta_{10}}{\alpha_{10}}) \delta \gamma_1^2)
\\ \nonumber & - &
P \sqrt{\alpha_{10} \beta_{10}} ( \frac{1}{8}(\frac{\delta \alpha_1}{\alpha_{10}} -
\frac{\delta \beta_1}{\beta_{10}})^2 + \frac{1}{2} \frac{\delta
\gamma_1^2}{\alpha_{10}\beta_{10}})
\\ & + &
\frac{2 \Omega}{\alpha_{10} \beta_{10}}((\delta \beta_2 - \delta \alpha_2) \delta \gamma_1
+ ( \delta \alpha_1 - \delta \beta_1) \delta \gamma_2)].
\eearr
For simplicity, first we set $ \Omega = 0 $.
Using the Euler-Lagrangian equation of motion, we can get the following two coupled equations 
of $ \delta \alpha_1 $ and $ \delta \beta_1 $.

\bearr
\delta \ddot \alpha_1 & = & \nonumber [\frac{P}{2}\alpha_{10}^2 \sqrt{ \frac{\beta_{10}}{\alpha_{10} }} - 4(1 + 
\epsilon) - 2 \lambda ( \frac{9}{\alpha_{10}} +  \frac{2}{\beta_{10}})] \delta \alpha_1
\\ & - & [\frac{P}{2} \alpha_{10}^2 \sqrt{\frac{ \alpha_{10}}{\beta_{10}}} + 2 \lambda 
\frac{\alpha_{10}}{\beta_{10}^2}], 
\delta \beta_1,
\eearr

\bearr
\delta \ddot \beta_1 & = & - [\frac{P}{2} \beta_{10}^2 \sqrt{\frac{\beta_{10}}{ \alpha_{10}}} + 2 \lambda
\frac{\beta_{10}}{\alpha_{10}^2}] \delta \alpha_1 
\\ \nonumber & + &  [\frac{P}{2}\beta_{10}^2 \sqrt{\frac{\alpha_{10}}{ \beta_{10}}} - 4(1 - \epsilon) - 
2 \lambda ( \frac{2}{\alpha_{10}} +  \frac{9}{\beta_{10}})] \delta \beta_1.
\eearr

For an isotropic trap, $ \epsilon = 0 $, and we set $ \delta \alpha_1 = - \delta \beta_1 $ to 
calculate the 
quadrupole mode ($ m_z \pm 2 $) frequency which is given by,
\be
\omega_q^2 = ( 4 + \frac{20 \lambda}{R_0} - P R_0^2 ),
\ee
where $ R_0 $ is the equilibrium radius of the system, and this  can be obtained
from the real solution of the cubic equation: $ (1+\frac{P}{2}) R_0^3 - R_0 - 4 \lambda = 0 $.
The above quadrupole frequency $ \omega_q $ is valid for all interaction (repulsive, attractive) strength
with a small anharmonic term ($\lambda $).
For $ \lambda = 0 $, the critical value of the interaction strength (above which the system is unstable) is
$ P = - 2 $.  
For $ \lambda \neq  0 $, and  $ P \leq -2 $, there is no real solution and the system collapses. Note that 
the quartic potential does
not affect the critical strength above which the system is unstable compared to the harmonic trapped
BEC.
The quadrupole mode frequency $ \omega_q $ vs. $ P $ is shown in the Fig.1.
\begin{figure}[h] \label{quadru}
\epsfxsize 9cm
\centerline {\epsfbox{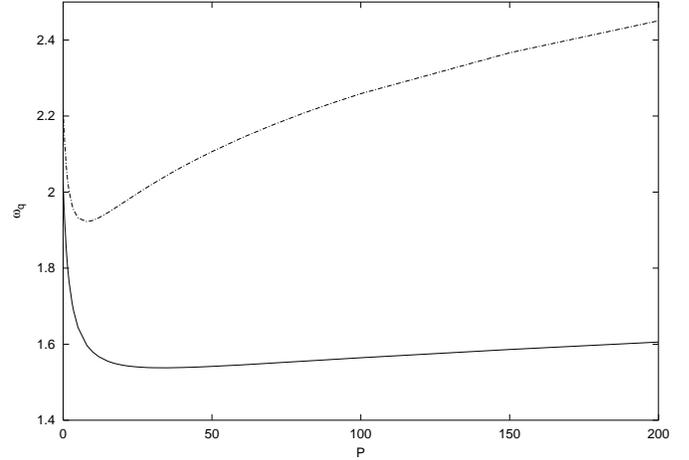}}
\vspace{0.5 cm}
\caption{Quadrupole mode frequency $\omega_q $ vs. interaction parameter $P$ for $ \lambda = 0.005 $ 
(solid) and $ \lambda = 0.05 $ (dashed-dot).}
\end{figure}
For $ \lambda = 0 = P $, it reproduces the known result for the
quadrupole mode frequency of a harmonically trapped non-interacting BEC.
Also, for $ \lambda = 0 $, it gives the known result of the quadrupole mode 
frequency of a harmonically trapped interacting BEC \cite{stringari}.
This result shows that the $ \omega_q $ is decreased compared  to the non-interacting case 
due to the effect of the repulsive interaction. Similarly, the $ \omega_q $ is also increased compared 
the non-interacting case due to the effect of the attractive interaction.
But small anisotropy ($ \lambda \neq 0 $) increases the $ \omega_q $ drastically in all cases.
Note that, the $ \omega_q $ depends on two parameters, $ \lambda $ and $ P $.
For a harmonically trapped BEC, $ \omega_q $ always decreases when $P$ increases, whereas 
for the additional quartic term, $ \omega_q $ first decreases for increasing $P$, but after some value of $P$, 
it starts increasing with $P$. The value of $P $ at which $\omega_q$ starts increasing depends on $ \lambda $. 
For example, $ P_c = 40 $ for $ \lambda = 0.005 $ and $ P_c = 10 $ for $ \lambda = 0.05 $.
   
The centrifugal term $ - \Omega L_z $ shifts the quadrupole mode frequency by $ \pm 2 \Omega $.
The lowest energy surface mode frequency is given by, 
\be
\omega_{-2} = \sqrt{ 4 + \frac{20 \lambda}{R_0} - P R_0^2} - 2 \Omega.
\ee

\section{ Rotating BEC confined in a harmonic trap}

Using the three coupled equations (\ref{c1}), (\ref{c2}) and (\ref{c3}) for the three variational 
parameters,
we get the following equation for the ratio of the widths:
\bearr \label{width}
 &  & (1+\epsilon -\Omega^2) \eta^4 + 2 (1 + \epsilon - \Omega^2)  \eta^3
+  2 \epsilon \eta^2
\\ \nonumber & +& 2 ( \Omega^2 - 1 + \epsilon ) \eta + (\Omega^2- 1 + \epsilon ) 
+ \frac{P}{2} \sqrt{\eta} [(1 + \epsilon - \Omega^2) \eta^3
\\ \nonumber & + & (1 + 3 \epsilon -5 \Omega^2) \eta^2 +  (5 \Omega^2 + 3 \epsilon
-1) \eta + (\Omega^2 - 1 + \epsilon )] = 0.
\eearr
The above relation for the aspect ratio of the system is valid for all interaction strength
including repulsive as well as attractive system also. The same kind of relation for the phase
parameter $ \gamma_2 $ is obtained in \cite{sinha}.

First, we consider a symmetric case ($\epsilon = 0 $) and $ P > 0 $.
One would expect naively that the ratio of the square of the widths, 
$ \eta $ becomes one for all rotation frequency. But this is not the case
for all rotation frequencies. $ \eta $ is always one when $ 0 \leq \Omega \leq \omega_q /2 $,
where $ \omega_q $ is the quadrupole mode frequency which is given by
\be
 \omega_{q}  = \sqrt{4 - P R_{0}^2}.
\ee
But $\eta$ has three solutions when $ \Omega $ is greater than $ \frac{\omega_q}{2} $.
One solution is still one and the other two solutions are less than one and greater than one.
But the system has the lowest energy when $\eta $ is either less than or greater than one. 
Now we discuss how the angular momentum is transferred to the system by the quadrupolar deformation.
When the rotation frequency is in the interval of $ 0 \leq \Omega \leq \omega_q /2 $, $\eta $ is always one 
and it is clear from Eq. (\ref{ang}) that the average angular momentum per particle is zero,
although we are rotating the system.
In other words, the system does not responds to the external rotation till $ \Omega 
=\frac{\omega_2}{2} $. This is a signature of the irrotationality of the condensate. 
Once the system starts deforming, it starts transferring angular momentum (see Eq. \ref{ang})
and vortex starts nucleate when the average angular momentum per particle is one. 
The existence of the critical angular frequency at which the angular momentum per particle suddenly 
goes up from $0$ to $1$ is the direct manifestation of the superfluid of the condensate. 
The average angular momentum of each particle has been measured experimentally which is one 
when a single vortex is nucleated \cite{mad1}.
One can do the same analysis for $ \epsilon \neq 0 $.
For large $P$ and very small $ \epsilon $, $\omega_{q} \sim \sqrt{2} $, then a single vortex 
will form when $\Omega \sim 1/\sqrt{2} \sim 0.7 $ which is experimentally observed.

For non-interacting system, set $P = 0$ in Eq. (\ref{width}), then
one can get the following quartic equation for $\eta$:
\bearr
& & \nonumber (1+\epsilon -\Omega^2) \eta^4 + 2 (1 + \epsilon - \Omega^2)  \eta^3
+ 2 \epsilon \eta^2  \\ & + & 2 ( \Omega^2 - 1 + \epsilon ) \eta
+ (\Omega^2 - 1 + \epsilon ) = 0.
\eearr

In the non-interacting system, $ \eta $ is complex if the rotating frequency
$ \Omega $ is in the interval [ $\sqrt{1-\epsilon}$,  $\sqrt{1+\epsilon}$].
Only one real solution exists for $ \Omega < \sqrt{1-\epsilon} $ and
$ \Omega > \sqrt{1+\epsilon} $ \cite{recati,sinha}. For symmetric trap potential ($\epsilon =0$) it is  
clear that there is only one real solution ($\eta = 1$) exist for any  rotating frequency $ \Omega $.
The lowest energy surface mode frequency is $ 2 \omega_0 $ and hence instability will
occur when $ \Omega = \omega_0 $ at which the system is destabilized.
There is no spontaneous shape deformation and hence one can not produce a vortex in the non-interacting 
system because the angular momentum per particle is always zero.
We have also checked that for finite but small $ \epsilon $, $ \eta $ changes very slowly with $ \Omega $ 
which is not sufficient
to transfer one unit of angular momentum per particle to the system and hence there is no vortex formation
in the non-interacting system even for finite $ \epsilon $.

Now we consider an attractive BEC. We will always consider $ |P| <2$, since the condensate will 
be in stable state if $ |P| < 2 $. In this attractive case, the quadrupole mode frequency is given below.
\be 
\omega_{q} = \sqrt{ 4 + \frac{2|P|}{2-|P|}}.
\ee
For any values of $|P| < 2 $, $ \omega_q > 2$. 
It clearly shows and confirmed by the solution of the Eq. (\ref{width}) that the shape deformation will not 
take 
place in the range of $ \Omega = 0 $ to $\Omega = 1 $. In other words, $\eta$ is always one. Since there is no 
shape deformation and consequently vortex nucleation is not possible. 
Note that authors in Ref.\cite{gunn1,mottelson} showed that the vortex formation in a rotating attractive BEC in
harmonic trap is prohibited due to the fact that all the angular momentum is carried by the center of mass
motion. 
Our analysis also shows why there is no vortex formation in an attractive BEC in harmonic trap. One can do the 
same analysis 
for $ \epsilon \neq 0 $, but the deformation is not enough to transfer unit quantum of angular momentum to each 
particle. 

\section{Rotating BEC confined in a quadratic-plus-quartic potential}
In this section we will discuss the effect of the quartic potential on the possibility of a vortex 
formation in a harmonically trapped BEC.
Using the three 
coupled equations (\ref{c1}), 
(\ref{c2}) and (\ref{c3}), we get the following two coupled polynomial equations:

\bearr \label{c11}
0 & = & \nonumber (\alpha_{10}^3 \beta_{10}^3 + 2 \alpha_{10}^4 \beta_{10}^2 +  \alpha_{10}^5 
\beta_{10})(1+\frac{P}{2} 
\sqrt{\frac{\beta_{10}}{\alpha_{10}}}) \\ \nonumber & - & \alpha_{10}^3 \beta_{10} ( 3 \Omega^2 + 1 + 
\epsilon ) + (\alpha_{10} \beta_{10}^3 + 2 \alpha_{10}^2 \beta_{10}^2)(\Omega^2 - 1 - \epsilon) 
\\ & - & \lambda (5 \alpha_{10}^2 
\beta_{10} + 7  \alpha_{10} \beta_{10}^2  +  \alpha_{10} + 3 \beta_{10}^3)
\eearr
 and
\bearr \label{c22}
 0 & = & \nonumber (\alpha_{10}^3 \beta_{10}^3 + 2 \alpha_{10}^2 \beta_{10}^4 +  \alpha_{10} 
\beta_{10}^5)(1+\frac{P}{2}
\sqrt{\frac{\alpha_{10}}{\beta_{10}}}) \\ \nonumber & - &  \alpha_{10} \beta_{10}^3 ( 3 \Omega^2 + 1 +
\epsilon ) + (\alpha_{10}^3 \beta_{10} + 2 \alpha_{10}^2 \beta_{10}^2)(\Omega^2 - 1 - \epsilon)
\\  & - &  \lambda (5 \alpha_{10} \beta_{10}^2 + 7  \alpha_{10}^2 \beta_{10} + 3 \alpha_{10}^3 + 
\beta_{10}^3).
\eearr
To show that there is a circular symmetry breaking of a BEC
confined in a quadratic-plus-quartic potential, we put $ \epsilon = 0 $ and consider
large $P$ (say $P = 1000$) and solve those two coupled equations.
We find that there is a circular symmetry breaking when the rotating frequency is greater
than one-half of the quadrupole mode frequency $ \omega_{q} $. 
We plot $\eta $ vs. $ \Omega $ for $ \epsilon = 0 $ and $ \lambda = 0.005 $ which is given in Fig.2.
In this case, $ \frac{\omega_q}{2} = 0.887179 $ and $\eta $ is always one for
$\Omega = 0 $ to $ \Omega = 0.887179 $.
When $ \Omega > 0.887179 $, $\eta$ has three real and positive solutions and $\eta <1 $ or $ \eta > 1 $ 
corresponds to the lowest energy state. As the system starts deforming itself, 
it transfers angular momentum to the system.
When $ \Omega = 0.905 $, each particle gets an unit angular momentum, i.e. $< L_z> = N \hbar $.
The vortex formation can occur for $ \Omega < 1 $ as well as $ \Omega > 1$, depending on the
parameter $ P $ and $ \lambda $. For $ \lambda = 0.05 $, the critical rotation frequencies are $ \Omega_c = 
1.2, 1.422, 1.554 $ for
$P = 100, 500, 1000 $, respectively. 
\begin{figure}[h]
\epsfxsize 9cm
\centerline {\epsfbox{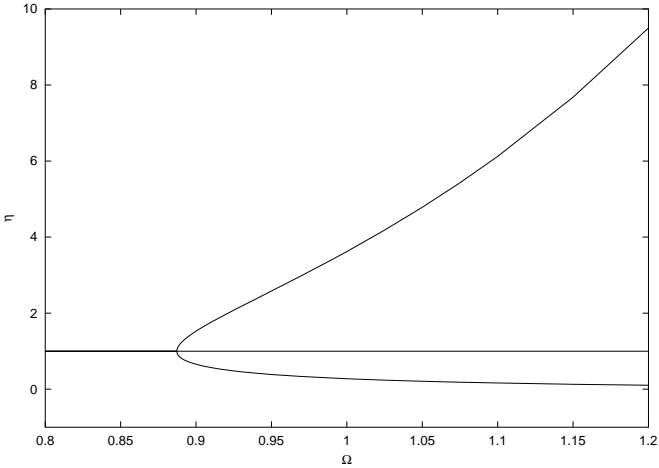}}
\vspace{0.5 cm}
\caption{The variation of $ \eta $ vs. $\Omega $ for $\lambda = 0.005 $ and large $P$.}
\end{figure}

The shape deformation also occurs for non-interacting and attractive Bose gases. 
To discuss about the shape deformation in non-interacting system, first we set $ P = 0$ in the equations 
(\ref{c11}) and (\ref{c22}). We find that the $ \eta $ has three  real and positive 
solutions when $ \Omega > \frac{\omega_q}{2} $, like in the condensate in the harmonic trap. 
For quantitative discussion, we take $ \lambda = 0.005 $, then the system
starts deformation when $ \Omega > \frac{\omega_q}{2} = 1.0123 $. In the previous section 
we have seen that the spontaneous deformation takes place 
due to the two-body interaction. But here spontaneous shape deformation exists 
in an ideal Bose system confined in a quadratic-plus-quartic potential. 
The average angular momentum per particle is always zero when $ \Omega  = 0 $ to $ \Omega = 1.0123 $, but
$ < L_z > = N \hbar $ at $ \Omega = 1.027 $. We give another example for $\lambda = 0.05 $. For this anharmonic 
strength,
the system will starts deforming when $\Omega > \frac{\omega_q}{2} = 1.10895 $ and the vortex will form when 
$ <L_z> = N \hbar $ at $\Omega = 1.21 $. It shows that as we increase the strength of the anharmonic term,
the critical rotational frequency is also increasing. 
Note that the vortex formation occurs only when $ \Omega > 1 $ because 
$ \frac{\omega_q}{2} $ is always greater than 1.
The variation of $\eta $ vs. $ \Omega $ for non-interacting BEC is shown in the Fig.3.
\begin{figure}[h]
\epsfxsize 9cm
\centerline {\epsfbox{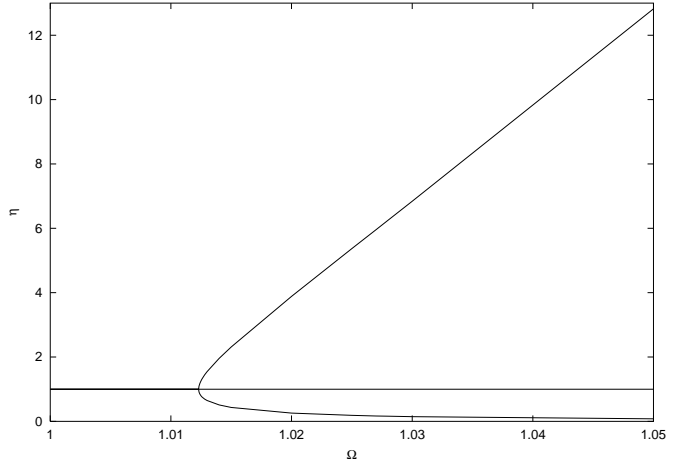}}
\vspace{0.5 cm}
\caption{The variation of $ \eta $ vs. $ \Omega $ for $P=0$, $ \lambda = 0.005 $ and $\epsilon = 0$.}
\end{figure}
It is worth to point out that the shape deformation can even occur 
in the non-interacting Bose gas if it is confined in a quadratic-plus-quartic potential.

In the previous section, we have seen that harmonically trapped attractive BEC 
can not hold a vortex. This is because of the absence of the shape deformation 
in the rotating system. Now we will see what happens if we add the quartic potential 
in Hamiltonian. Once we add the quartic potential, we can
access a large rotation frequency ($\Omega \geq 1$) for which the rotating attractive system is not 
destabilized.
We assume $ P = -0.015 $ and $ \lambda =0.01 $ and then we get $ \frac{\omega_q}{2} = 1.02606 $ for the 
symmetric trap.
By solving the equations (\ref{c11}) and (\ref{c22}), we see that the spontaneous shape deformation starts 
when
$\Omega \sim 1.02606 $ and transferring an angular momentum to the system. Each particle gets an unit angular 
momentum, $ <L_z> = N \hbar $, when $ \Omega = 1.052 $.
We give another example. Assume, $ P = -1.5 $ and $ \lambda =0.05 $ and then we get 
$ \frac{\omega_q}{2} = 1.66215 $ for the symmetric trap. By solving the equations (\ref{c11}) and (\ref{c22}),
we get three real and positive solution of $\eta$ when $\Omega > \frac{\omega_q}{2}$.
But the existence of the three solutions of $\eta $ does not mean that the system will starts deforming itself.
Actually, the lowest energy state can be obtained only for
$\eta =1$. It means there is no shape deformation for these parameters and hence a vortex can not appear 
under this conditions.
In this attractive system, the shape deformation can occur which leads to the vortex nucleation, 
depending on the strengths of the two-body interaction and the quartic potential.
When there is no shape deformation for a given large $P$ and $\lambda$,
the center of mass motion is excited \cite{gunn1,mottelson}.
These results are also consistent with the recent results of Lundh {\em et al.} \cite{lundh1}.
Note that $\frac{\omega_q}{2} $ is always greater than 1, so that deformation takes place only 
when $ \Omega > 1 $. 

\section{Summary and Conclusions}
In this paper we have calculated the quadrupole mode frequency. 
We have shown that when the rotating frequency of an interacting BEC confined in the 
quadratic-plus-quartic 
potential is greater than one-half of the quadrupole mode frequency, the system itself starts
deforming from a circular shape to an elliptic shape and hence it breaks
the rotational symmetry of the Hamiltonian.
We have shown how this self-deformation transfer an angular momentum to the rotating system and 
consequently creates a vortex. 

We have argued that the formation of the vortex is not possible due to the absence of 
the spontaneous shape deformation of a harmonically trapped rotating attractive BEC as well as an ideal BEC.
But the shape deformation and consequently the vortex formation can occur in a rotating attractive BEC if it 
is confined by an additional quartic potential. 
We have also shown that the shape deformation occurs not only due to the two-body repulsive interaction,
it can occur even in the non-interacting as well as attractive BEC if it is confined in a 
quadratic-plus-quartic potential and consequently it produces a vortex.
However, the shape deformation and consequently forming a quantized 
vortex in a rotating attractive BEC depends on the strengths of the two-body interaction and the 
quartic potential.

Note that, in actual experiments the lowest energy surface mode is excited through 
the rotating potential with small anisotropies. In this work we have studied (for simplicity) 
the vortex formation in a symmetrically trapped BEC, but we can easily extend our work with small
anisotropy $\epsilon$.

\end{multicols}
\end{document}